\documentclass[twocolumn,showpacs,preprintnumbers,amsmath,amssymb,widetext,super
scriptaddress]{revtex4}

\usepackage{graphicx}

\begin{document}

\title{Coulomb-Blockade directional coupler}
\author{P. Pingue}
\email{pingue@sns.it}
\author{V. Piazza}
\author{F. Beltram}
\affiliation{NEST-INFM \& Scuola Normale Superiore, I-56126 Pisa, Italy}
\author{I. Farrer}
\author{D.A. Ritchie}
\author{M. Pepper}
\affiliation{Cavendish Laboratory, University of Cambridge,
Madingley Road, Cambridge CB3 0HE, United Kingdom}

\begin{abstract}
A tunable directional coupler based on Coulomb Blockade 
effect is presented. Two electron waveguides are coupled by a
quantum dot to an injector waveguide.
Electron confinement is obtained by
surface Schottky gates on single GaAs/AlGaAs heterojunction.
Magneto-electrical measurements down to 350 mK are presented and
large transconductance oscillations are reported on both outputs
up to 4.2 K. Experimental results are interpreted in terms of
Coulomb Blockade effect and the relevance of the present design
strategy for the implementation of an electronic multiplexer is underlined.

\end{abstract}
\pacs{73.23.Ad, 73.23.Hk, 73.63.Nm}
\maketitle

Controlled directional injection of electrons from one electronic waveguide to
another is being intensively investigated owing to its importance in
wavelength multiplexing and in telecommunication routing devices.
While photonic multiplexers represent a mature technology,
electronic steering devices are still at their infancy.

Pioneering designs were demonstrated exploiting a field-effect
tunable barrier between two semiconductor waveguides
\cite{delAlamo1,delAlamo2,delAlamo3,Tsukada1,Tsukada2}, 
implementing a Y-branch switch
or switching the electrons by means of electric
side-gates\cite{Y-branch1,Y-branch2,Y-branch3}. An electronic
device based on two electron waveguides coupled by an open interaction 
window rather than a tunneling barrier was also
proposed\cite{Wang,Vanbesien,Takagaki} and realized\cite{Hirayama},
but so far no switching behavior was demonstrated in 
this configuration. 

In this Letter we describe a scheme where the coupling element
between two semiconductor waveguides is a single quantum dot (QD)
and Coulomb blockade (CB) governs electron routing. We shall
demonstrate that in the classic configuration with two electron
waveguides mixing at an open ballistic window a QD can be induced --
with appropriate biasing conditions and geometry -- and
employed as a gate-controlled coupling element between the two waveguides.

\begin{figure}[ht!]
\begin{center}
\includegraphics[width=7 cm,clip]{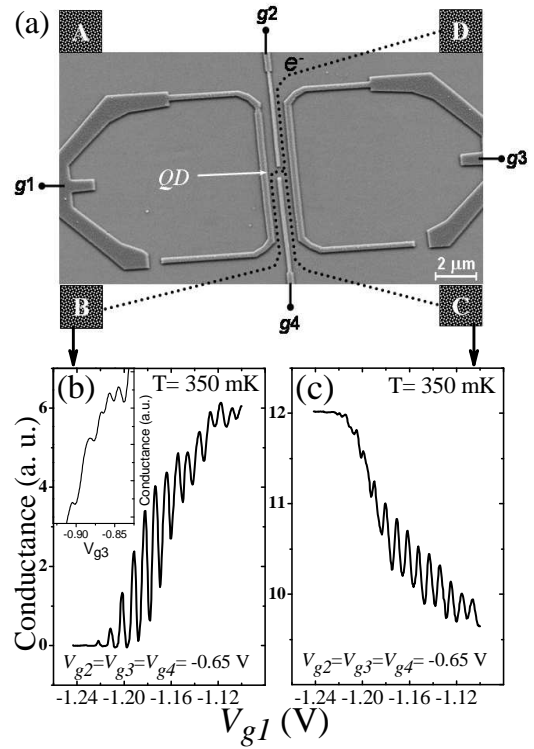}
\end{center}
\caption{(a) SEM picture of the coupled-waveguide device. 
Schottky gates are labelled as {\sl g1} through {\sl g4}.
Letters {\sl A} to {\sl D} represent the Ohmic contacts 
(not shown); (b) differential conductance in the collector B
(trough the QD) and (c) corresponding conductance in the collector
C. Inset: "symmetric dot" conductance in function of $V_{g3}$.}
\label{fig1}
\end{figure}

Figure~\ref{fig1}(a) shows
a scanning electron microscope (SEM) picture of the device. Schottky gates are labeled
{\sl g1} through {\sl g4}.
The Al split-gate structure was nanofabricated by e-beam lithography and 
thermal evaporation. The waveguides of the device described in this work have
a total length $L=7.5$~$\mu$m each and a geometrical width $W=0.44$~$\mu$m. The
central split-gate is $w=0.17$~$\mu$m wide while the coupling window is $l=0.57$~$\mu$m long. The GaAs/AlGaAs heterostructure contains a two-dimensional
electron gas (2DEG) located 70 nm below the surface
with a mobility of $\mu$~=~3.7$\times 10^5$ cm$^{2}$/Vs and a
carrier concentration of {\it n}~=~2.4$\times$10$^{11}$cm$^{-2}$ at
350 mK.

In our experiment gates {\sl g2} to {\sl g4} were negatively
biased by a DC voltage ($V_{g2}, V_{g3}, V_{g4}$, respectively),
while gate {\sl g1} (biased $V_{g1}$) was employed as a plunger gate.
A drain-source AC voltage (V$_{DS}$=100 $\mu$V) was applied at
the input waveguide (Ohmic contact D in Fig.~1).

The AC current flowing through outputs B and C was measured by
two current preamplifiers and phase-locked techniques.
A lower AC excitation of 20 $\mu$V was also used to test that the
device was in the linear regime also at the lowest temperatures.

\begin{figure}[ht!]
\begin{center}
\includegraphics[width=8.0 cm,clip]{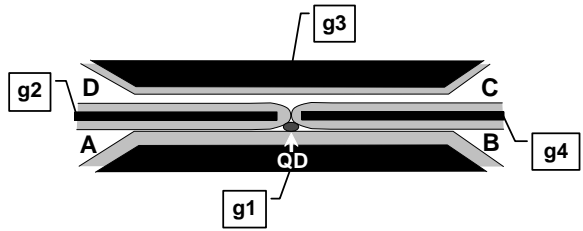}
\end{center}
\caption{Schematic of our device. QD indicates the quantum-dot location
(black circle). Black regions represent the metal gates while gray areas
show the regions depleted by the applied bias.}
\label{fig2}
\end{figure}

Panels (B) and (C) in Fig.~1 show the differential conductance
measurements relative to collectors B and C respectively when
an AC bias is applied to contact D. The Ohmic-contact A
was left floating. Biasing $V_{g2}$, $V_{g3}$, and $V_{g4}$ at -0.65 V, high-contrast
oscillations in the output current appear on both collector
waveguides at 350 mK as a function of $V_{g1}$ in the range from
-1.24 V to -1.1 V.
One of the outputs (Fig.~1, C) displays a $5\%$-wide
current modulation while the other collector (Fig.1, B) shows
a $100\%$-wide current modulation oscillation in the output current.

This behavior can be explained taking into account the formation
of a QD in the CB regime in the region
indicated by the black disk in the schematic sketch shown in Fig.~\ref{fig2}. 
The QD presence in that position was verified in all
cool-downs, and a specific characterization of each waveguide was
performed in order to exclude the presence of unintentional dots
in the input or output channels\cite{vanHouten,Staring}.
By symmetrically reversing the biasing configuration of gate electrodes a dot symmetrically located in the opposite of the coupling region could be induced. This dot was indeed
observed when contact B was employed as emitter and A and D as
collector waveguides (A, D). In the case of the device shown here it yielded a lower contrast
in conductance measurements (see inset in Fig.~\ref{fig1}B). We
attribute this behavior to a different coupling regime to the
reservoir, probably related to the tunnel barrier of the output
channel. 

The CB-oscillation pattern obtained by measuring the current from
contacts B and C showed a $\pi$ phase shift, excluding the possibility that
the QD extends to the whole central region and demonstrating the
switching behavior of our device.

Figure~\ref{fig2} shows a scheme of our device:
gate-depleted regions in the 2DEG are indicated by gray areas.
It is quite intuitive that a confined dot can be created when the coupling
window is pinched-off and when the bottom waveguide is almost
closed. Tunnel barriers originate from small grains or
lithographic imperfections in the metallic gates
that induce a constriction between the QD and the adjacent waveguides.

\begin{figure}[ht!]
\begin{center}
\includegraphics[width=8.5 cm,clip]{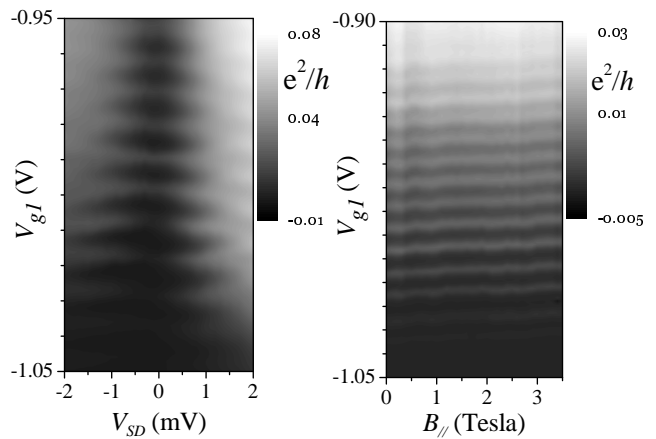}
\end{center}
\caption{Conductance behavior of the QD at finite bias(a)
and in magnetic field parallel to the 2DEG(b). T~=~350 mK.}
\label{fig3}
\end{figure}

In the following, a low-temperature study of the magnetotransport
properties of the dot represented in the scheme of Fig.~\ref{fig2}
is reported. CB characterization was performed as a
function of source-drain bias ($V_{SD}$) and as a function of a
magnetic field parallel to the 2DEG plane, in order to minimize
orbital magnetic-coupling effects. Figure~\ref{fig3}(a) shows the well-known
Coulomb diamonds in the QD conductance as a function of 
$V_{SD}$ and $V_{g1}$. The height
(in the $V_{SD}$ direction) of the diamonds can be used to measure the
charging energy $U_{CB}$ between two adjacent electron levels\cite{Kouwenhoven}.
From the maximum Coulomb gap $U_{CB}$={\it
e}$^{2}$/C$_{TOT}\simeq$1.6 meV a total dot capacitance of
$C_{TOT}\simeq$~100 aF is deduced.

In a simple model of an ungated 2D disk-shaped QD the total
capacitance is given by $C_{TOT}=4\epsilon_{0}\epsilon_{r}D$, where $D$ is the QD diameter,
and $\epsilon_{r}$=12.5 is the dielectric constant of GaAs. From
the measured value of $C_{TOT}$=100 aF we derive a dot diameter of D~=~0.23~$\mu$m,
in good agreement with the geometry of our device.
From this diameter and the charge density of electrons in
the original 2DEG, we estimate that there are $\sim$100 electrons in the
dot under the operating conditions of Fig.\ref{fig3}(a).

The mean periodicity of the Coulomb oscillations reported in
Fig.~1 $\Delta V_{g1}={\it
e}/C_{g1}$=9.4 mV corresponds to a gate capacitance of $C_{g1}$=
17 aF giving, therefore, $\eta$=C$_{g1}$/C$_{TOT}$= 0.17 as
``lever arm'' value between the applied gate voltage and the change in
the total energy of the island. The period $\Delta V_{g1}$
remains almost constant changing $V_{g1}$ down to the pinch-off
and no contribution related to discrete energy levels of the dot
is observed. In any case, the pinch-off in our device is
determined by that relative to the output channel QD-B, so no
information about the occupation number of the dot is directly
available through these measurements.

More information about the QD position can be extracted by comparing CB
oscillations as a function of $V_{g1}$ at different voltages applied
to the remaining gates. We observed an increase of $C_{g1}$
from 16.3 aF to 19 aF while decreasing $V_{g2}$, $V_{g3}$ and $V_{g2}$
from -0.6 V to -0.8 V, indicating that electrons in the dot
are pushed toward {\sl g1} when the other gates are biased with
increasingly negative voltages.

\begin{figure}[ht!]
\begin{center}
\includegraphics[width=8 cm,clip]{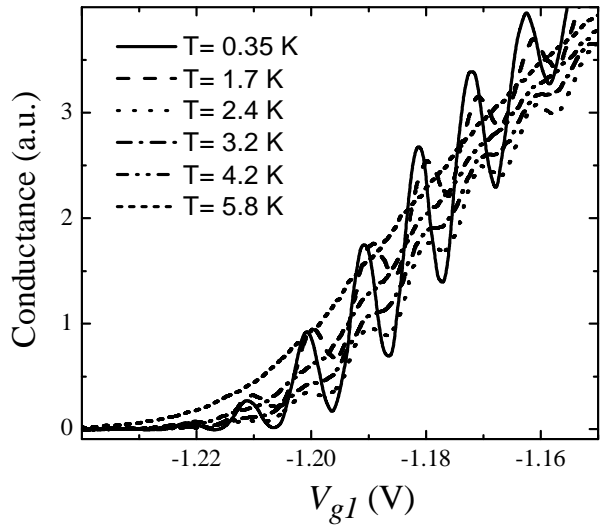}
\end{center}
\caption{Temperature behavior of the CB effect in the collector B:
conductance oscillations are present from 0.35 K up to 4.2 K.}
\label{fig4}
\end{figure}

Measurements in parallel magnetic field $B_{\parallel}$(Fig.\ref{fig3}(b))
show a common shift of the CB peaks that is caused by the coupling between the
magnetic field itself and the transverse components of electron
wave functions in the dot and in the leads (diamagnetic shift)\cite{Weis,Duncan}.
By plotting the peak spacing in order to eliminate common peak motion
with magnetic field and to minimize the presence of switching noise,
a strong fluctuation in the data is observed (data not shown).
The peak spacing does not follow a clear linear behavior with the magnetic
field as {\it g}$\mu_{B}B_{\parallel}/\eta$ (where {\it g} is the Land\'{e} factor
for bulk GaAs, $\mu_{B}$ the Bohr magneton and $\eta$ the lever arm value)
as expected on the basis of Zeeman splitting of the dot levels. This irregular
behavior has been already observed in gate-depleted QDs in the "weak-coupling"
regime\cite{Lindemann,Folk}: the QD is indeed weakly coupled to 
drain and source and CB peaks typically have a height lower than
0.1 {\it e}$^{2}$/${\it h}$ (see Fig.\ref{fig3}(b)) and an
irregular pattern of the peak spacing in function of the magnetic
field is reported.

Finally, the temperature behavior of our device (see
Fig.\ref{fig4}) shows that the effect is still present up to 4.2
K, consistently with the CB charging energy $U_{CB}$ previously found.

We wish to point out an advantage intrinsic to this scheme with
respect to quantum devices based on coherence effects: 
as for the case of single electron transistors 
\cite{Fulton,Shirakashi}, appropriate geometries and materials can lead 
to higher and even room-temperature operation of such CB-based directional coupler. 
Available transconductance values are very high. In fact at low temperatures switching voltages are
in the mV range and at least one order of
magnitude lower than those required in Y-branch devices 
(operated both in the so-called "external side-gate"\cite{Y-branch2} and 
in "internal ballistic" switching mode\cite{Y-branch3}). 
This leads to measured normalized-transconductance
values as high as $I^{-1}dI/dV_{g1}\sim1500$ V$^{-1}$ for collector B
in the present implementation. 
This characteristic is of interest in terms of low power consumption
for high miniaturization and large scale integration.

In conclusion, a directional coupler device based on CB was
demonstrated. Its differential-conductance characterization
as a function of magnetic field and temperature was presented. In the 
CB regime this device behaves like a current switch
in one collector output and as a current modulator in the opposite
one. The same basic scheme can be employed to design logic functions
 or, employing a series of our device, a CB-based electronic
multiplexer.

Fruitful discussions with S. de Franceschi and 
M. Governale are gratefully acknowledged. This work was supported 
in part by the European Research and Training Network COLLECT 
(Project HPRN-CT-2002-00291). Work at NEST-INFM was supported 
in part by MIUR under FIRB contract RBNE01FSWY, 
and work at Cavendish Laboratory by EPSRC.

\end{document}